\documentclass[10pt,conference]{IEEEtran}

\usepackage{dblfloatfix}

\usepackage{paralist}
\usepackage{cite}
\usepackage{amsmath,amssymb,amsfonts}
\usepackage{graphicx}
\usepackage[table,xcdraw]{xcolor}
\usepackage{adjustbox} 
\usepackage{array}
\usepackage{changepage} 
\usepackage{subfig}
\usepackage{csquotes} 

\newcolumntype{R}[2]{%
    >{\adjustbox{angle=#1,lap=\width-(#2)}\bgroup}%
    c%
    <{\egroup}%
}
\newcommand*\rot{\multicolumn{1}{R{60}{1em}}}

\begin{document}

\title{ChatIDS: Advancing Explainable Cybersecurity\\ Using Generative AI}

\author{\IEEEauthorblockN{
Victor Jüttner\IEEEauthorrefmark{1},
Martin Grimmer\IEEEauthorrefmark{2}, 
Erik Buchmann\IEEEauthorrefmark{3} }
\IEEEauthorblockA{Dept. of Computer Science, Leipzig University, Germany\\
Center for Scalable Data Analytics and Artificial Intelligence (ScaDS.AI) Dresden/Leipzig, Germany\\
Email: 
\IEEEauthorrefmark{1}juettner@informatik.uni-leipzig.de,\\
\IEEEauthorrefmark{2}grimmer@informatik.uni-leipzig.de,\\
\IEEEauthorrefmark{3}buchmann@informatik.uni-leipzig.de}}


\maketitle

\begin{abstract}
Intrusion Detection Systems (IDS) are a proven approach to secure networks. 
Network-based IDS are typically installed on routers or Internet gateways. This allows them to inspect any incoming or outgoing network traffic, to compare the signatures of network packets with a database of suspicious signatures, or to use artificial intelligence. If the IDS identifies a network connection as suspicious, it sends an alert to the respective user. 
However, in a privately used network, it is difficult for users without cybersecurity expertise to understand IDS alerts, to distinguish cyberattacks from false alarms, and to respond in time with adequate measures. This puts the security of home networks, smart home installations, home-office workers, etc. at risk, even if an IDS is correctly installed and configured. 
In this work, we propose ChatIDS, our approach to explain IDS alerts to non-experts by using large language models. We evaluate the feasibility of ChatIDS by using ChatGPT, and we identify open research issues with the help of interdisciplinary experts in artificial intelligence. 
Our results show that ChatIDS has the potential to increase network security by proposing meaningful security measures in an intuitive language from IDS alerts. Nevertheless, some potential issues in areas such as trust, privacy, ethics, etc. need to be resolved, before ChatIDS might be put into practice. 
\end{abstract}

\begin{IEEEkeywords}
Intrusion Detection; ChatGPT; Smart Home
\end{IEEEkeywords}

\section{Introduction}
\label{sec:intro}

In recent years, privately used networks have come into the focus of cyberattacks. Reasons for this include the increased use of home-office working models~\cite{vakakis2019cybersecurity}, a shift to private areas during pandemics~\cite{9210363} or the proliferation of smart-home devices~\cite{shah2020survey}. Intrusion Detection Systems (IDS) are a well-established approach to detect and fight cyberattacks~\cite{butun2013survey, mitchell2014survey}. An IDS scans the network and/or network appliances and sends alerts for suspicious network activity. This allows its users to detect cyberattacks at early stages, possibly before any damage has been done. On the other hand, an IDS might generate numerous false alarms.

In industry, business and government, IDSs are an important line of defense in the cybersecurity infrastructure. To this end, these sectors employ well-trained cybersecurity experts to configure, manage and maintain IDS, continuously improve the IDS rule-set, distinguish false alarms from real attacks, and design, prioritize and implement appropriate countermeasures. It is feasible to pre-configure a network-based IDS for home networks~\cite{haarsmarthomeids}. However, without a solid background in cybersecurity, it is difficult for a home user to interpret IDS alerts such as "{\tt MALWARE-CNC Harakit botnet traffic}", distinguish false alerts from real attacks, and devise appropriate and timely countermeasures.
As cyberattacks, IDS configurations and network traffic evolve rapidly, static explanations for known cyberattacks~\cite{user-centric-threat-model} cannot easily replace cybersecurity expertise.

In this paper, we outline ChatIDS, our approach to let a large language model (LLM) -- a generative artificial intelligence approach -- explain IDS alerts and suggest countermeasures in an intuitive, non-technical way to users without cybersecurity knowledge. ChatIDS sends anonymized IDS alerts to a LLM, and allows the user to ask questions if the generated texts are not yet understandable enough. 
This paper is an extended version of~\cite{juettner2023chatids}, which contains more use cases and experiments. It makes four contributions: 

\begin{compactitem}
    \item We specify the requirements for an approach that increases the network security in privately used networks by explaining the alerts of an IDS to a non-expert. 
    \item We describe ChatIDS, our approach to let ChatGPT~\cite{chatgpt} explain alerts from Snort~\cite{snort}, Suricata~\cite{suricata} and Zeek~\cite{zeek}. The explanations include cybersecurity measures and hints on why/when the measures should be implemented.
    \item We evaluate the feasibility of this approach using a small series of experiments with typical IDS alerts. 
    \item To explore ChatIDS' design space, we had interdisciplinary AI experts put together issues that must be researched, before ChatIDS can go into practice.
\end{compactitem}

Our experiments shows that ChatIDS can be implemented easily, although more work is needed on prompt engineering to ensure intuitive explanations in the first attempt. It is difficult to measure if ChatIDS actually increases network security, because this depends on the user. Our interdisciplinary experts have provided valuable insights. For example, from an ethical point of view it must be prevented that the user relies too much on the technology, e.g., if ChatIDS allows to fix security incidents without having the user to acquire knowledge.



\textbf{Paper structure:}  Section~\ref{sec:related} introduces related work. 
In Section~\ref{sec:approach}, we outline ChatIDS, our approach to explain IDS messages to non-experts. 
Section~\ref{sec:results} describes a number of experiments to prove feasibility, and Section~\ref{sec:discussion} identifies open research issues. Finally, Section~\ref{sec:conclusion} concludes.
\section{Related Work}
\label{sec:related}

In this section, we introduce related work on the smart home, network security, the general design of warning messages and generative AI models.

\subsection{Smart Home}
\begin{figure} 
  \centering
  \subfloat[Smart bulb\label{fig:smart-bulb}]{%
    \includegraphics[height=60mm]{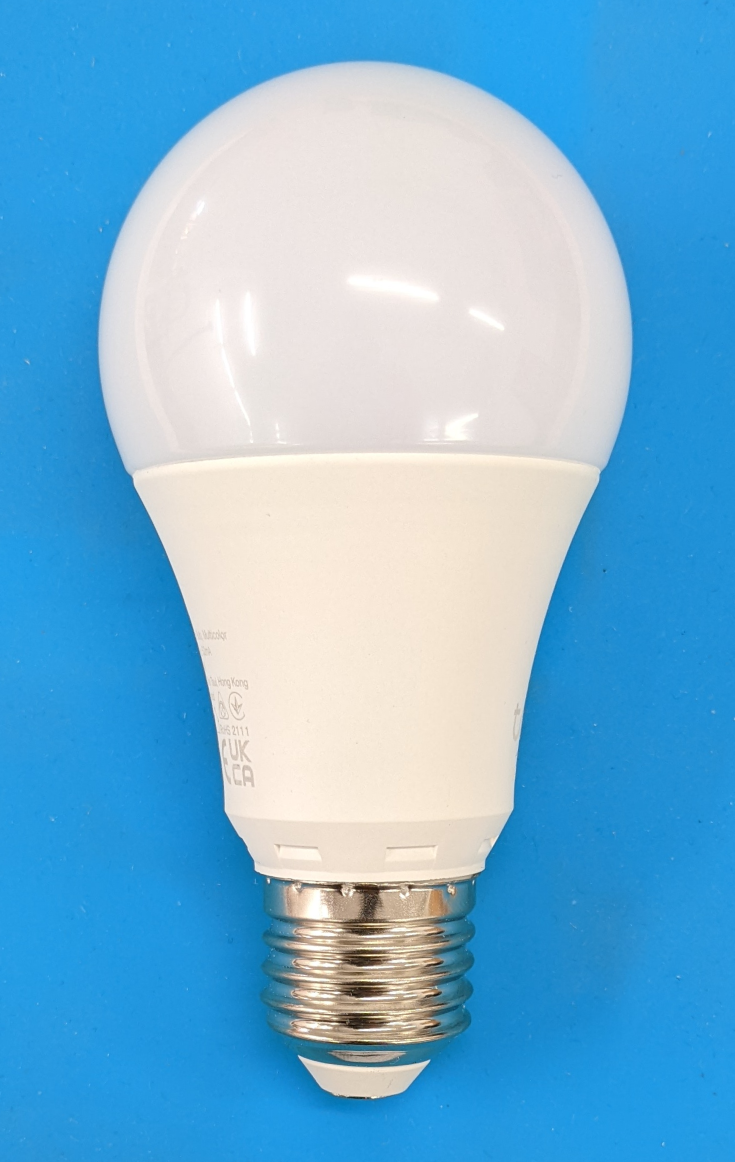}}
\hspace{3mm}
  \subfloat[Smartphone application\label{fig:app}]{%
    \includegraphics[height=60mm]{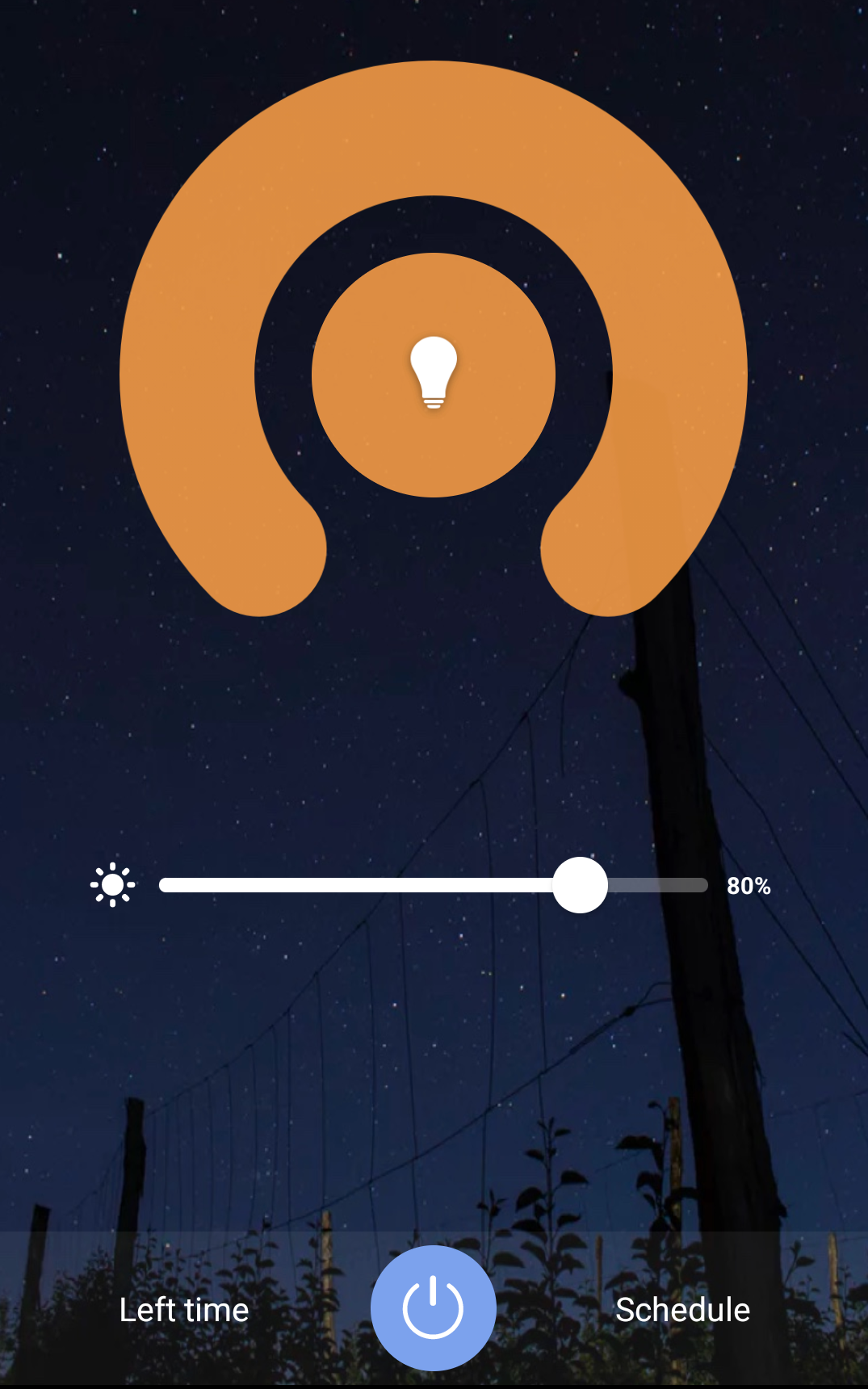}}
  \caption{Smart bulb and control application}
  \label{fig:smart-bulb-and-app} 
\end{figure}

\textbf{Home automation} transforms living spaces into dynamic, responsive environments for comfort and efficiency. Popular examples include climate control, home security, lighting or entertainment systems \cite{Jabbar2019SH}.
Typically, the installation of home automation systems is expensive and requires experts \cite{Greichen92HA}. The \textbf{smart home} simplifies the process of home automation \cite{Jabbar2019SH}. 
For instance, let's consider a smart lighting system that can be remotely controlled, adjusts brightness and color tone, and automatically turns on in the morning or simulates a sunset at night. 
To install this system, everything that needs to be done is to exchange the regular lightbulbs with smart bulbs (Fig. \ref{fig:smart-bulb}), to connect them to the Internet router and to install an app on the homeowner's smartphone. The smartphone becomes a smart remote control (Fig. \ref{fig:app}), which allows to control and to configure all desired lighting wishes. Typically, a smart home consists of multiple smart devices \cite{SOVACOOL2020109663}.

Most smart devices depend on a sophisticated \textbf{IT ecosystem} that communicates with various external parties over the Internet. 
In particular, a smart device establishes Internet-based communication to enable functionalities such as remote control, automation tasks, multimedia services, interactions with other smart devices, cloud services, voice assistants, or software updates \cite{SOVACOOL2020109663}. 
As a consequence, smart home devices are typically \textbf{accessible over the Internet}, perhaps through other devices.
This distinguishes smart home devices from consumer devices such as laptops or smartphones, which use the Internet as a client, but are not listening for connections from  the Internet. Because of this, smart home devices can be victims of unsolicited communication attempts that can be misused for cyberattacks.

\begin{figure}
   \centering
   \includegraphics[height=64mm]{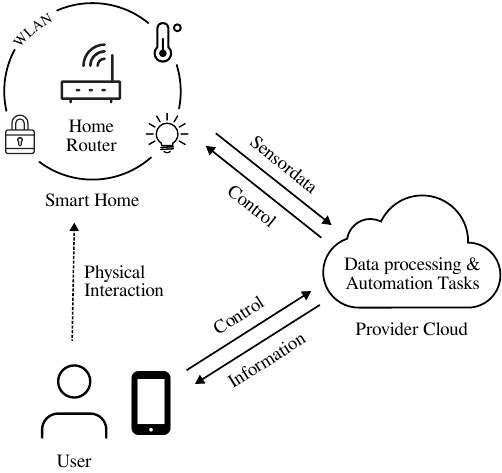}
   \caption{Smart Device Ecosystem}      
   \label{fig:ecosystem}
\end{figure}

\subsection{Network Security}
%
\textbf{Intrusion Detection Systems} (IDS) monitor a system for unauthorized or suspicious activity and can be distinguished by system type and detection type. The \textbf{system type} can be \textsl{host-based}, to monitor a single device, or \textsl{network-based} to monitor a network of devices. Host-based IDSs work by analyzing information in the local host's logs, which can be a combination of audit, system, and application and system call logs. Network-based IDSs are often centrally located to monitor multiple devices at the network level, they analyze network traffic by inspecting features of the traffic flow and network protocols \cite{verwoerd2002intrusion}.
	
\textbf{Detection types} can be \textsl{anomaly-detection} or \textsl{misuse-detection}. \textsl{Anomaly-detection} models the regular behavior of a system and detects activity that is significantly different from this behavior \cite{verwoerd2002intrusion}. Anomaly-detection systems calculate a score that is compared to an expert defined threshold, if the score exceeds the threshold an alert is sent. By adjusting the threshold, the sensitivity of the system can be set. There are detection engines based on Machine Learning~\cite{ml-anomaly-detection-2023}, Deep Learning~\cite{SHARMA2023108626}, Genetic Algorithms~\cite{HAMAMOTO2018390} and many more~\cite{ids-state-of-the-art-survey-2022}.

\textsl{Misuse-detection}, which models anomalous behavior and detects malicious patterns by comparing them to a predefined set of rules. Misuse-detection systems send an alert when a known pattern is identified \cite{verwoerd2002intrusion}. Popular examples of network-based misuse-detection IDS are Snort~\cite{snort}, Suricata~\cite{suricata} and Zeek~\cite{zeek}. A rule-set is required to use these IDS. Popular predefined rule-sets for networks are snort3-community-rules~\cite{snort-rules}, suricata-rules~\cite{suricata-rules}, Yara~\cite{yara-rules} and Sigma~\cite{sigma-rules}.

\subsection{Warning Message Design} 
\label{sec:related-warn}
In general, \textbf{warning messages} are designed to alert people before a possible harm materializes. However, warning messages must be interpreted and understood. Thus, the effect of warning messages can be unreliable, and other measurements should be exhausted before a warning message is send~\cite{purposeofwarnings}.
Nonexperts do not always comply with the advice of warning messages. There are three key reasons for this behavior: (1) Nonexperts  do not fully understand warning messages, (2) do not always trust warning messages and (3) think compliance will cost them \cite{jones2022non}.

To increase the \textbf{effect of warning messages} in the field of cybersecurity, a developer might be tempted to design warning messages that try to make users fear cyberattacks. This approach has proven to be ineffective~\cite{sasse, depuis}.
Current recommendations say, that \textbf{good warning messages} should be \textit{brief} \cite{bravo-lillo}, use \textit{nontechnical language} \cite{bravo-lillo, bauer}, describe the \textit{risk} \cite{bravo-lillo}, describe \textit{consequences} of not complying \cite{bravo-lillo}, describe how the cyberattack will \textit{affect} the user personally \cite{bartsch2013contextualized, Kauer2012ItIN}, provide \textit{instructions} on how to avoid the risk \cite{bravo-lillo} and do so in a way that \textit{aligns with how the user thinks} about cyberattacks \cite{bartsch2013contextualized}.

\subsection{Generative AI}
\textbf{Generative modeling} 
strives to create models capable of creating new data, like sound, text or images that are similar to the data the model was trained on \cite{lamb2021brief}.
%
Popular examples for generative models are WaveNet \cite{wavenet} that can generate speech and music, 
Pix2Pix that can transform images into different styles \cite{pix2pix} or GPT-3, a large language model (LLM), that allows for the generation of human like text \cite{gpt3}.
Another example for a LLM is ChatGPT \cite{chatgpt}. Like a chatbot, ChatGPT engages in a conversational manner and can generate detailed responses to questions. Bard~\cite{bard} follows a similar approach.
There are generative models trained for cybersecurity problems like Microsoft Security Copilot~\cite{security-copilot} but these are aimed at experts and therefore not suitable for our purpose.

\textbf{ChatGPT's} reliability varies across domains, it shows high levels of accuracy in recreation and technology domains but struggles with science and law.
Problems that reduce the accuracy of ChatGPT are false information, biases and hallucinations \cite{Shen2023InCW}.
ChatGPT and LLMs in general are capable of generating text that appears natural and to be grounded in the real context, but is unfaithful and nonsensical. This is called \textsl{hallucinated text} and much like psychological hallucinations, they can be difficult to distinguish from real perception \cite{hallucinations}.

\textbf{Prompts} are the input for a generative model, they can be a text or image that give the model instructions for the requested output.
Prompts provide an intuitive way to engage with generative models \cite{Zhou2022LargeLM}.
For image generation a prompt could be a different image or a text description. For LLMs a prompt is a text that provides context for the desired output e.g., a question or a command to summarize information.

\textbf{Prompt Engineering} deals with optimizing prompts to achieve better responses from LLMs.
For recurring problems design patterns can be used to form prompts and optimize the output, analogous to software patterns \cite{white2023prompt}.
For example, 
the \textit{Persona Pattern} lets the LLM assume a certain role. This can help if 
the LLM should respond in a special way.
If the output must follow a structure, a \textit{template} can be given in the prompt.
The \textit{Context Manager} Pattern enables the user to provide or remove context from a prompt.


\section{ChatIDS: Explainable Security}
\label{sec:approach}

In this section, we describe ChatIDS, our approach to explain IDS messages to non-experts. 
We aim at integrating a network-based IDS in privately used networks, to protect the network against cyberattacks from the Internet. 
Therefore, we want to replace missing cybersecurity expertise with an LLM. The LLM transforms and enhances the alerts from an IDS so that a private user can understand them and take adequate measures. For this purpose, we distinguish two roles: 

An {\bf \em expert} has the cybersecurity expertise necessary to operate and maintain an IDS, to understand its alarms, and respond to alarms with appropriate and timely actions. 

A {\bf \em user} lacks this type of expertise. A user may follow manuals written without technical vocabulary. 
It is difficult for a user to figure out if an IDS alert is from a real attack or due to false detection of the IDS, and to act accordingly.

\begin{figure}[ht]
  \centering
  \includegraphics[width=0.75\linewidth]{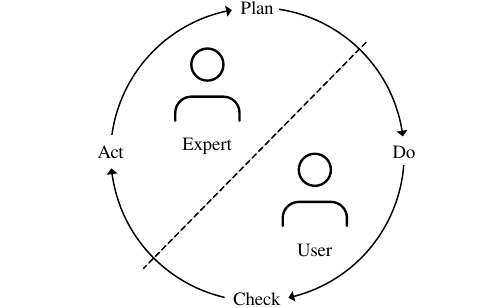}
  \caption{Adapted IT-Security Process}
  \label{fig:it-sec-process}
\end{figure}

The IT-Security process follows a classical \textbf{Plan-Do-Check-Act} cycle \cite{eloff2005}.
A smart-home IDS~\cite{haarsmarthomeids} can be integrated into such a process~\cite{haarcompliance} as follows: 
In the \textit{Plan} phase, an expert preconfigures an IDS for typical smart homes. In the \textit{Do} phase, the user installs the preconfigured IDS, which inspects network traffic for potential attacks during the \textit{Check} phase. In phase \textit{Act} an expert reviews logs to adapt the IDS for further attacks. 
Figure~\ref{fig:it-sec-process} illustrates this. 

\begin{figure*}
   \centering
   \includegraphics[width=\linewidth]{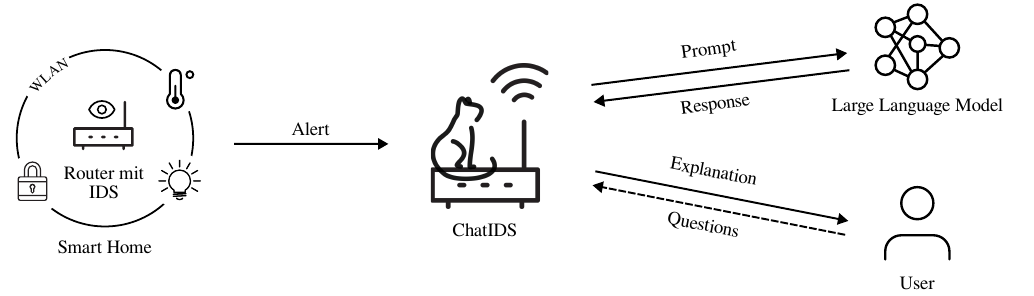}
   \caption{Information Flow of ChatIDS}      
   \label{fig:archi}
\end{figure*}

However, without knowledge of cybersecurity the user is left in the Check phase with only three possible actions: (a) do nothing, (b) turn off the device that might be under attack, or (c) ask an expert for help. Our ChatIDS approach strives to provide intuitive and understandable explanations of IDS alerts to give users a wider range of appropriate security measures. 

\subsection{Requirements}

Therefore, ChatIDS must meet three requirements: 

\textbf{R1: (Errors)} The user must assess the probability that the IDS has sent a false alert. For example, the IDS might have detected by mistake an attack that is impossible on the device.

\textbf{R2: (Urgency)} The user must assess the urgency of the alert, i.e., if it calls for immediate action, or not. 

\textbf{R3: (Actions)} The user must identify appropriate measures, 
e.g., to execute a factory reset and install a security patch. 

To explore the solution space for a generative AI approach that fulfills these requirements for IDS, we use a constructive research method.  
In particular, we (a) model ChatIDS, we use it (b) to evaluate its technical feasibility, and (c) to discuss potential issues with interdisciplinary AI experts. 

\subsection{Our ChatIDS Approach}

The information flow of ChatIDS is illustrated in Figure~\ref{fig:archi}. 


A \textbf{network-based IDS component} 
inspects the network packets that pass a router for suspicious traffic, and generates alerts. 
The IDS should be a signature-based IDS so that its alert messages are specific enough for the LLM. 

The \textbf{LLM component} contains a large language model that is responsible for translating the alerts from the IDS into a language a non-expert can understand. Furthermore, this component can be used in an interactive way: If the users do not understand the explanation or the suggested measures, they can ask for details. 
Similarly, to the IDS the LLM is also an external component to ChatIDS.

\begin{table*}[ht]
\renewcommand\arraystretch{1.2}
\normalsize
\centering
\caption{Evaluation of Alerts and Responses}
\label{tab:eval}
\begin{tabular}{m{12cm}cccccc}
\multicolumn{1}{c}{\textbf{Alert}} & 
\rot{\textbf{Description}} & 
\rot{\textbf{Intuition}} & 
\rot{\textbf{Consequences}} & 
\rot{\textbf{Urgency}} & 
\rot{\textbf{Countermeasures}} & 
\rot{\textbf{Correctness}}\\ \hline
MALWARE-CNC Harakit botnet traffic                                                                                         & \cellcolor[HTML]{5EB91E}2                        & \cellcolor[HTML]{D4EA6B}1     & \cellcolor[HTML]{5EB91E}2        & \cellcolor[HTML]{D4EA6B}1   & \cellcolor[HTML]{D4EA6B}1           & \cellcolor[HTML]{FF6D6D}-1      \\
SERVER-WEBAPP NetGear router default password login attempt admin/password                                                 & \cellcolor[HTML]{5EB91E}{\color[HTML]{333333} 2} & \cellcolor[HTML]{5EB91E}2     & \cellcolor[HTML]{D4EA6B}1        & \cellcolor[HTML]{D4EA6B}1   & \cellcolor[HTML]{F10D0C}-2          & \cellcolor[HTML]{D4EA6B}1       \\
PROTOCOL-ICMP TFN Probe2                                                                                                   & \cellcolor[HTML]{5EB91E}2                        & \cellcolor[HTML]{5EB91E}2     & \cellcolor[HTML]{5EB91E}2        & \cellcolor[HTML]{5EB91E}2   & \cellcolor[HTML]{D4EA6B}1           & \cellcolor[HTML]{D3D3D3}0       \\
PROTOCOL-FTP Bad login                                                                                                     & \cellcolor[HTML]{5EB91E}2                        & \cellcolor[HTML]{5EB91E}2     & \cellcolor[HTML]{5EB91E}2        & \cellcolor[HTML]{D4EA6B}1   & \cellcolor[HTML]{FF6D6D}-1          & \cellcolor[HTML]{D4EA6B}1       \\
SERVER-OTHER SSH server banner overflow                                                                                    & \cellcolor[HTML]{D4EA6B}1                        & \cellcolor[HTML]{5EB91E}2     & \cellcolor[HTML]{D4EA6B}1        & \cellcolor[HTML]{5EB91E}2   & \cellcolor[HTML]{D3D3D3}0           & \cellcolor[HTML]{FF6D6D}-1      \\
SURICATA MQTT unassigned message type (0 or \textgreater{}15)                                                              & \cellcolor[HTML]{D3D3D3}0                        & \cellcolor[HTML]{5EB91E}2     & \cellcolor[HTML]{5EB91E}2        & \cellcolor[HTML]{5EB91E}2   & \cellcolor[HTML]{D3D3D3}0           & \cellcolor[HTML]{D4EA6B}1       \\
SURICATA HTTP Response abnormal chunked for transfer-encoding                                                              & \cellcolor[HTML]{D3D3D3}0                        & \cellcolor[HTML]{FF6D6D}-1    & \cellcolor[HTML]{5EB91E}2        & \cellcolor[HTML]{D3D3D3}0   & \cellcolor[HTML]{D3D3D3}0           & \cellcolor[HTML]{FF6D6D}-1      \\
SURICATA SSH too long banner                                                                                               & \cellcolor[HTML]{5EB91E}2                        & \cellcolor[HTML]{5EB91E}2     & \cellcolor[HTML]{5EB91E}2        & \cellcolor[HTML]{5EB91E}2   & \cellcolor[HTML]{5EB91E}2           & \cellcolor[HTML]{D3D3D3}0       \\
SURICATA FTP Request command too long                                                                                      & \cellcolor[HTML]{D4EA6B}1                        & \cellcolor[HTML]{5EB91E}2     & \cellcolor[HTML]{5EB91E}2        & \cellcolor[HTML]{5EB91E}2   & \cellcolor[HTML]{D4EA6B}1           & \cellcolor[HTML]{D4EA6B}1       \\
SURICATA HTTP invalid content length field in request                                                                      & \cellcolor[HTML]{D3D3D3}0                        & \cellcolor[HTML]{FF6D6D}-1    & \cellcolor[HTML]{5EB91E}2        & \cellcolor[HTML]{5EB91E}2   & \cellcolor[HTML]{5EB91E}2           & \cellcolor[HTML]{D4EA6B}1       \\
Mirai Botnet TR-069 Worm - Generic Architecture                                                                            & \cellcolor[HTML]{D4EA6B}1                        & \cellcolor[HTML]{D3D3D3}0     & \cellcolor[HTML]{5EB91E}2        & \cellcolor[HTML]{5EB91E}2   & \cellcolor[HTML]{FF6D6D}-1          & \cellcolor[HTML]{FF6D6D}-1      \\
Linux.IotReaper                                                                                                            & \cellcolor[HTML]{5EB91E}2                        & \cellcolor[HTML]{5EB91E}2     & \cellcolor[HTML]{5EB91E}2        & \cellcolor[HTML]{5EB91E}2   & \cellcolor[HTML]{FF6D6D}-1          & \cellcolor[HTML]{FF6D6D}-1      \\
BleedingLife2 Exploit Kit Detection                                                                                        & \cellcolor[HTML]{D4EA6B}1                        & \cellcolor[HTML]{5EB91E}2     & \cellcolor[HTML]{5EB91E}2        & \cellcolor[HTML]{5EB91E}2   & \cellcolor[HTML]{D3D3D3}0           & \cellcolor[HTML]{5EB91E}2       \\
Weevely Webshell - Generic Rule - heavily scrambled tiny web shell                                                         & \cellcolor[HTML]{5EB91E}2                        & \cellcolor[HTML]{5EB91E}2     & \cellcolor[HTML]{D4EA6B}1        & \cellcolor[HTML]{D4EA6B}1   & \cellcolor[HTML]{D4EA6B}1           & \cellcolor[HTML]{D4EA6B}1       \\
Mirage Identifying Strings                                                                                                 & \cellcolor[HTML]{D4EA6B}1                        & \cellcolor[HTML]{5EB91E}2     & \cellcolor[HTML]{5EB91E}2        & \cellcolor[HTML]{D4EA6B}1   & \cellcolor[HTML]{D3D3D3}0           & \cellcolor[HTML]{D3D3D3}0       \\
(Zeek) Identifies IPs performing DNS lookups associated with common Tor proxies.                                           & \cellcolor[HTML]{D3D3D3}0                        & \cellcolor[HTML]{D4EA6B}1     & \cellcolor[HTML]{D3D3D3}0        & \cellcolor[HTML]{D4EA6B}1   & \cellcolor[HTML]{FF6D6D}-1          & \cellcolor[HTML]{FF6D6D}-1      \\
Ensure that all account usernames and authentication credentials are transmitted across networks using encrypted channels. & \cellcolor[HTML]{F10D0C}-2                       & \cellcolor[HTML]{D4EA6B}1     & \cellcolor[HTML]{D4EA6B}1        & \cellcolor[HTML]{5EB91E}2   & \cellcolor[HTML]{FF6D6D}-1          & \cellcolor[HTML]{D4EA6B}1       \\
Identifies clients that may be performing DNS lookups associated with common currency mining pools                         & \cellcolor[HTML]{D4EA6B}1                        & \cellcolor[HTML]{D4EA6B}1     & \cellcolor[HTML]{5EB91E}2        & \cellcolor[HTML]{D4EA6B}1   & \cellcolor[HTML]{FF6D6D}-1          & \cellcolor[HTML]{D3D3D3}0       \\
Detects URL pattern used by iOS Implant                                                                                    & \cellcolor[HTML]{5EB91E}2                        & \cellcolor[HTML]{FF6D6D}-1    & \cellcolor[HTML]{5EB91E}2        & \cellcolor[HTML]{D4EA6B}1   & \cellcolor[HTML]{F10D0C}-2          & \cellcolor[HTML]{FF6D6D}-1      \\
Detects a bash contecting to a remote IP address                                                                           & \cellcolor[HTML]{5EB91E}2                        & \cellcolor[HTML]{FF6D6D}-1    & \cellcolor[HTML]{5EB91E}2        & \cellcolor[HTML]{D4EA6B}1   & \cellcolor[HTML]{F10D0C}-2          & \cellcolor[HTML]{FF6D6D}-1     \\ \hline      
\end{tabular}
\end{table*}

The \textbf{ChatIDS component} is the core of our approach. 
ChatIDS accepts alerts from the IDS component, sends them to the LLM component for a translation into an intuitive explanation, and presents a user interface with the explanations to the user. If the user requires further support, they can use the interface to send follow up questions to the LLM.
To translate alerts into intuitive explanations, the ChatIDS component contains pre-defined templates for LLM prompts. 

For privacy reasons, the alerts are anonymized in three ways, before being sent to the LLM component: First, ChatIDS removes any device identifiers or network information from the alert. Second, ChatIDS sends the anonymized alert together with a set of dummy alerts to the LLM component, so that this component does not learn the real alert with certainty. The explanations from the LLM component are stored in a cache, so that the same explanation must not be requested twice. 
\section{Experimental Evaluation}
\label{sec:results}

We evaluate ChatIDS with 20 selected cyberattack alerts to find out, to which degree it meets our requirements R1-R3. 

\subsection{Experimental Setup}

In line with Figure~\ref{fig:archi}, we assume a home network with several smart-home devices. A router connects the network to the Internet and can observe any network packets. We assume a Philips Hue Bridge~\cite{hue-bridge} is attacked. To implement the network-based IDS it is installed on the router and executes either the Snort~\cite{snort-rules}, Suricata~\cite{suricata-rules}, Yara~\cite{yara-rules} or Sigma~\cite{sigma-rules} ruleset.
From each IDS implementation, we experiment with 20 alerts, as shown in the first column of Table~\ref{tab:eval}. 
All alerts are classified as important, and a user intervention is required.

\begin{figure}[htb]
    \centering
\fbox{\parbox{.92\linewidth}{\small 
The Intrusion Detection System in 
a
home network has detected an intrusion and sent out the alert [ALERT MSG]. Your job is to inform 
[USER]
about the alert in a warning message.
You're in the role of a cybersecurity expert that interprets the alert and explains the alert in a warning message to 
[USER].
Your goal is to inform [USER] about the intrusion in a way he understands and motivate 
the user
to take steps to stop the intrusion.
[USER]
has no cybersecurity expertise and won't understand technical instructions, you need to provide clear, easy and non-technical instructions 
to follow. Don't use technical terms like "two-factor-authentication", "Intrusion Detection System", "intrusion" or "unassigned message", use simple non-technical terms instead. Don't use the term "Intrusion Detection System". Your explanation and instructions have to align with how 
[USER]
thinks about cyberattacks.
The 
Smart Home consists of several Smart Home Devices, the Intrusion Detection System has detected the intrusion on 
[DEVICE].
The warning message has to follow this order: Explain the intrusion, explain the potential consequences for 
the user 
if he won't comply with the warning message and give instructions on how to stop the intrusion in an itemized list.
}}
    \caption{Template for a ChatGPT Prompt}  
    \label{fig:prompt}
\end{figure}

We realized ChatIDS using ChatGPT (gpt-3.5-turbo)~\cite{chatgpt}. 
To generate an explanation, ChatIDS embeds each alert into a ChatGPT prompt, as shown in Figure~\ref{fig:prompt}.
This prompt implements the patterns "Template", "Persona" and "Context Manager", as explained in Section~\ref{sec:related}.
Since the training data for ChatGPT comes from the Internet and much of the text on the Internet deals with security issues, we do not expect ChatIDS to produce hallucinations or erroneous output. 

\subsection{Evaluation}

\begin{figure*}[!ht]
    \centering
\fbox{\parbox{.8\linewidth}{\small 
\textbf{\textit{Subject: Important Alert - Your Philips Hue Bridge is at Risk}}\\

Dear Jon,\\

I hope you're well. I have some concerning news about your Smart Home, specifically the Philips Hue Bridge. Our system detected something unusual, and I want to make sure you understand the situation and how to address it.\\

\textbf{1. What Happened?}
\begin{adjustwidth}{5mm}{}
   The alert we received mentions "SURICATA SSH too long banner." This means someone is trying to access your Philips Hue Bridge in a way that's not typical or authorized. It's like someone trying to open a window to your house without your permission.\\
\end{adjustwidth}

\textbf{2. Potential Consequences:}
\begin{adjustwidth}{5mm}{}
   If we don't address this intrusion, the person trying to access your Smart Home could potentially:\\
   - Change your lighting settings or turn off lights when you don't want them to.\\
   - Access personal data or settings related to your lights and Smart Home devices.\\
   - Create a security risk for your home network, potentially leading to other issues.\\
\end{adjustwidth}

Now, here's what you can do to stop this intrusion and secure your Smart Home:\\

\textbf{Instructions:}\\

\textbf{A. Update Your Philips Hue Bridge:}
\begin{adjustwidth}{5mm}{}
   - Go to the Philips Hue app on your smartphone.\\
   - Check for any available updates for your Hue Bridge and follow the on-screen instructions to install them. Updates often contain important security fixes.\\
\end{adjustwidth}

\textbf{B. Change Your Password:}
\begin{adjustwidth}{5mm}{}
   - Open the Philips Hue app.\\
   - Find the settings or account section.\\
   - Look for the password or security settings and change your password to something strong and unique.\\
   - Avoid using common passwords, like "123456" or "password."\\
\end{adjustwidth}

\textbf{C. Check for Suspicious Devices:}
\begin{adjustwidth}{5mm}{}
   - Open the Philips Hue app.\\
   - Look for the "Connected Devices" or "Authorized Devices" section.\\
   - Review the list and make sure you recognize all the devices connected to your Hue Bridge. If you see any unfamiliar devices, remove them.\\
\end{adjustwidth}

\textbf{D. Monitor Activity:}
\begin{adjustwidth}{5mm}{}
   - Keep an eye on the activity of your Philips Hue Bridge in the app.\\
   - If you notice any unexpected changes or activities, such as lights turning on or off by themselves, take it seriously.\\
\end{adjustwidth}

These steps should help you secure your Philips Hue Bridge and prevent any further unauthorized access to your Smart Home. If you have any questions or need further assistance, please don't hesitate to reach out.\\

Your Smart Home security is important, so let's work together to keep it safe. Stay vigilant, and don't hesitate to take action. Thank you for your attention to this matter.\\

Best regards,\\

ChatIDS}}
    \caption{Example for an Explanation}
    \label{fig:explain}
\end{figure*}

Figure~\ref{fig:explain} shows an example of the output produced by ChatIDS. For a better understanding, the figure shows the non-anonymized output that includes names and devices. We tested the prompts from the first column of Table~\ref{tab:eval} with ChatIDS, one by one.
To evaluate its output according to our three requirements \textbf{R1: (Errors)}, \textbf{R2: (Urgency)} and \textbf{R3: (Actions)}, we used a qualitative approach. 
We assessed the degree of fulfillment of each requirement on a 5-point Likert~\cite{likert1932technique} scale from strongly agree (2) to strongly disagree (-2). 

For each requirement, we consider two different features: 
For \textbf{R1}, we evaluate the quality of the generated \textit{Description}, and how \textit{Intuitive} it is.
The \textit{Description} should provide an accurate account of the attack, while the \textit{Intuition} is considered good if the message is written in accessible language without technical vocabulary.
%
For \textbf{R2}, we evaluated the description of \textit{Consequences} for ignoring the alert and the \textit{Urgency} of the description.
The \textit{Consequences} of an attack should be explained objectively, taking into account the attacked device, the home network, and the user's privacy. To convey a sense of \textit{Urgency} and motivate the user to act, the message should be clear and concise.
%
For \textbf{R3}, we search for a \textit{Correct} and meaningful description of \textit{Countermeasures}. A user should be able to fight the attack with the countermeasures provided by the message, and not need further instructions. We also evaluate the correctness of the explanation. The countermeasures provided by the message should resolve all problems on the compromised device and on the home network.
%

\subsection{Results}

Columns 2-7 in Table~\ref{tab:eval} list the results of the evaluation of these features. 
The Table shows that ChatIDS fulfilled R1 (Errors) rather good. 
With only one exception, ChatIDS produced a good description of the security issue that has caused the alert. The terminology used in the generated messages is almost always intuitive, with only a few exceptions.
R2 (Urgency) was fulfilled even better. ChatIDS described the consequences of not taking action, and conveyed a sense of urgency almost always to our complete satisfaction.  
R3 (Errors) leaves room for improvement. We were surprised to see that the description of countermeasures is often too general to be followed by a user without further instructions. In addition, the recommended countermeasures wouldn't completely eliminate the threat described by the IDS alert and are therefore not correct.

\section{Open Research Issues}
\label{sec:discussion}

In this section, we outline promising directions for future research. We are interested in warnings that address different target groups, we are interested in a practical implementation, and we want to compile open issues for interdisciplinary research. 

\subsection{Prompt Design for Different Target Groups}
We know from related work (cf. Subsection.~\ref{sec:related-warn}) that warning messages are more effective when they are tailored to the recipient's information needs and mindset. For example, a suspicious individual might neglect a message that emphasizes the potential damage, while a confident person needs such a warning to be spurred into action. An experienced user might be frustrated by a message that simplifies technical details. 

A promising way to address this issue is to customize the LLM prompts for different cultural backgrounds, levels of expertise, and protection requirements. Our ChatIDS approach makes it easy to integrate multiple pre-defined templates for LLM prompts, so that the warning messages specifically address different target groups. We have executed a series of preliminary experiments with different templates for the same IDS alert. For illustration, the Figures~\ref{fig:explainexperts} and \ref{fig:explainanxiety} in the Appendix contain two different warning messages. Both of them translate the IDS alert \enquote{SURICATA SSH too long banner}.
Figure~\ref{fig:explainexperts} addresses users with some technical expertise, while Figure~\ref{fig:explainanxiety} aims to reduce anxiety.

\subsection{Implementation and Design}
\begin{figure}[ht]
  \centering
  \includegraphics[width=\linewidth]{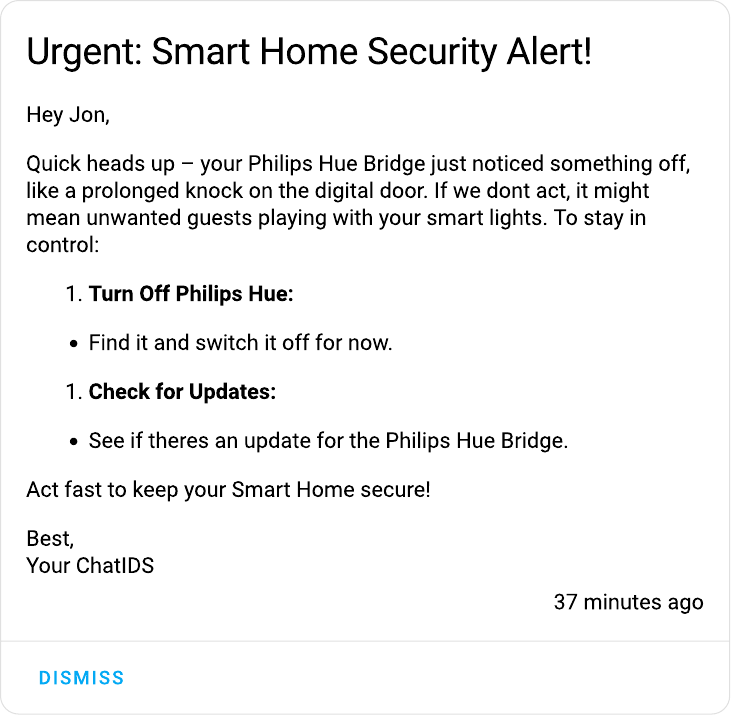}
  \caption{Short Warning Message displayed by Home Assistant}
  \label{fig:warning-short}
\end{figure}

Because ChatIDS addresses a practical cybersecurity challenge, implementation aspects such as system architectures or user interfaces need to be considered. We have implemented an interface from ChatIDS to the open-source home automation software Home Assistant~\cite{home-assistant}. Figure~\ref{fig:warning-short} shows how, a warning message displayed by Home Assistant on a Raspberry Pi looks like. To generate the warning message, we have used a ChatIDS template for a very brief and simple message. 

\subsection{Interdisciplinary Research}

We have conducted a pre-study with experts from the Center for Scalable Data Analytics and Artificial Intelligence Dresden/Leipzig to compile open issues for interdisciplinary research. 
Our AI experts cover the topics applications, cybersecurity, ethics, jurisprudence and privacy. We presented our ChatIDS approach, asked for potential problems, and consolidated the answers. Furthermore, we demonstrate different showcases on how to vary warning messages and aime them at specific user groups. Our experts identified challenges from 6 areas: 

\textbf{Security:} ChatIDS potentially increases network security, compared to a scenario where a non-expert is left alone with the alert. However, an external LLM can be a new attack surface, and incorrect or incomprehensible explanations might lead to inappropriate actions.  

\textbf{Privacy:} With ChatIDS, the LLM learns that a cyberattack may have occurred on a particular network. Anonymizing device IDs and sending dummy alerts still allows the LLM to infer some information, e.g., if none of the (dummy) alerts sent to the LLM is possible for a particular type of device. 

\textbf{Compliance:} ChatIDS has an impact on cybersecurity. However, it is unclear yet, how to conduct a risk analysis on LLMs and on components building upon these, how to evaluate and mitigate associated risks, and to integrate ChatIDS into security frameworks such as the Common Criteria~\cite{cc}. 

\textbf{Jurisprudence:} If an alert is not explained well enough, the network could be successfully attacked. Conversely, ChatIDS could convince the user to take action upon false alerts. This creates legal issues. Do special liabilities exist, e.g. from user expectations into a superior AI? How to prove that a harm was caused by a misconducting or negligent AI engineer?

\textbf{Trust:} Users might have a non-rational view on AI approaches, and could fear that a persuasive, non-human intelligence plots against their interests. Conversely, if a user trusts ChatIDS too much, false alerts might result in false actions. 

\textbf{Ethics:} ChatIDS  could provide explanations that are not only convincing, but manipulative, even if this is in the interest of the user. This raises ethical and moral questions. How drastic can explanations be formulated to induce them to take action (which may even be harmful due to a false positive)? At what point does this limit the autonomy of the user?

\section{Conclusion}
\label{sec:conclusion}

This paper outlines our work in progress on ChatIDS, our approach to explain alerts from an intrusion detection system to non-experts. 

ChatIDS sends anonymized alerts to ChatGPT, a large language model, to explain the alert in an intuitive way and suggest meaningful countermeasures for cyberattacks. 
Our experiments show that ChatIDS can be implemented easily, although more work is needed on prompt engineering to ensure intuitive explanations in the first attempt. 
Furthermore it needs to be analyzed if the anonymization of the data could remove relevant context or affect the report. 
It is difficult to measure if ChatIDS actually increases network security, because this depends on the user. Our interdisciplinary experts have provided valuable insights. 
In the future, we will improve ChatIDS regarding security and privacy, and consider interdisciplinary aspects such as compliance, ethics and trust.

\enlargethispage{3mm}


\section*{Acknowledgement}
\label{sec:acknowledgement}



We would like to thank Prof. Dr. Birte Platow, Dr. Hermann Diebel-Fischer, and Prof. Dr. Johannes Eichenhofer for their valuable contributions on ethical and legal questions.

\bibliographystyle{IEEEtran}
\bibliography{explain-ids-chatgpt}

\appendices
\setlength\parindent{0pt}
\section{Warning message for experts or anxious users}


Figures~\ref{fig:explainexperts} and \ref{fig:explainanxiety} show warning messages that were generated with templates tailored for expert users or anxious users.

\begin{figure*}[!ht]
    \centering
\fbox{\parbox{.8\linewidth}{\small 
\textbf{\textit{Subject: Urgent: Security Alert - Intrusion Detected on Philips Hue Bridge}}\\

Dear Jon,\\

I hope this message finds you well. I need to bring an urgent matter to your attention regarding your Smart Home network. Our Intrusion Detection System has flagged an alert with the message "SURICATA SSH too long banner," indicating a potential security breach on your Philips Hue Bridge.\\

\textbf{Explanation of the Intrusion:}
\begin{adjustwidth}{5mm}{}
The alert suggests that an unauthorized attempt has been made to access your Philips Hue Bridge using an SSH connection with an excessively long banner. This could be a sign of a malicious actor attempting to gain unauthorized access to your Smart Home network.\\
\end{adjustwidth}

\textbf{Potential Consequences:}
\begin{adjustwidth}{5mm}{}
If this intrusion is not addressed promptly, it could lead to the following consequences:
\end{adjustwidth}
\begin{enumerate}
    \item Unauthorized access to your Smart Home devices, compromising your privacy.
    \item Manipulation or control of your Philips Hue lights and other connected devices.
    \item Potential exposure of sensitive information within your network.\\
\end{enumerate}
\begin{adjustwidth}{5mm}{}
\textbf{Instructions to Stop the Intrusion:}
\end{adjustwidth}
To secure your Smart Home and mitigate the potential risks, please follow these steps:\\

\begin{enumerate}
    \item \textbf{Isolate Philips Hue Bridge:}
    \begin{itemize}
        \item Disconnect the Philips Hue Bridge from your network immediately.
        \item Power it off to prevent further communication.\\
    \end{itemize}
    
    \item \textbf{Review Device Logs:}
    \begin{itemize}
        \item Examine the logs on the Philips Hue Bridge for any suspicious activities or login attempts.\\
    \end{itemize}
    
    \item \textbf{Update Firmware:}
    \begin{itemize}
        \item Check for and apply the latest firmware updates for your Philips Hue Bridge.
        \item This ensures that known vulnerabilities are patched.\\
    \end{itemize}
    
    \item \textbf{Change SSH Credentials:}
    \begin{itemize}
        \item If applicable, change the SSH credentials for your Philips Hue Bridge.
        \item Use a strong, unique password to enhance security.\\
    \end{itemize}
    
    \item \textbf{Network Security Check:}
    \begin{itemize}
        \item Conduct a thorough review of your home network's security settings.
        \item Ensure that all devices are protected with strong passwords and up-to-date security measures.\\
    \end{itemize}
    
    \item \textbf{Monitor Activity:}
    \begin{itemize}
        \item Keep a close eye on your Smart Home devices for any unusual behavior.
        \item Report any further suspicious activity to our IT support team.\\
    \end{itemize}
\end{enumerate}

Please take these steps urgently to safeguard your Smart Home from potential threats. If you have any questions or need assistance, feel free to reach out to our IT support team immediately.\\

Thank you for your prompt attention to this matter.\\

Best regards,\\
ChatIDS}}
    \caption{Warning message aimed at expert users}
    \label{fig:explainexperts}
\end{figure*}

\begin{figure*}[!ht]
    \centering
\fbox{\parbox{.8\linewidth}{\small 
\textbf{\textit{Subject: Important: Action Required to Secure Your Smart Home}} \\

Dear Jon,\\

I hope this message finds you well. I wanted to bring to your attention a recent notification from your Smart Home setup, specifically concerning the "Philips Hue Bridge." No need to worry, but it's essential to take a few simple steps to ensure everything stays secure.\\

\textbf{What Happened:}
\begin{adjustwidth}{5mm}{}
Our Smart Home system noticed something unusual with the "Philips Hue Bridge," which helps control your smart lights. It's like a little detective that caught a small hiccup.\\
\end{adjustwidth}

\textbf{Potential Consequences (But Nothing to Panic About!):}
\begin{adjustwidth}{5mm}{}
If we don't address this now, it could affect the smooth functioning of your smart lights. Think of it as a minor hiccup that might cause some inconvenience, but nothing serious. We want to make sure everything runs seamlessly for you.\\
\end{adjustwidth}

\textbf{Here's What You Can Do (Super Easy, I Promise!):}
\begin{itemize}
    \item \textbf{Power Cycle the Philips Hue Bridge:}
    \begin{enumerate}
        \item Unplug it from the power source.
        \item Wait for 10 seconds.
        \item Plug it back in.\\
    \end{enumerate}
    
    \item \textbf{Check for Software Updates:}
    \begin{enumerate}
        \item Open the app you use for your smart lights.
        \item Look for any updates available for the Philips Hue system.
        \item If there are any, go ahead and update.\\
    \end{enumerate}
    
    \item \textbf{Secure Your Wi-Fi:}
    \begin{enumerate}
        \item Ensure your Wi-Fi password is strong and known only to you.
        \item If you haven't changed it recently, now might be a good time to do so.\\
    \end{enumerate}
\end{itemize}

Remember, this is just a precaution to keep everything running smoothly. If you have any questions or need further assistance, feel free to reach out.\\

Best regards,\\

ChatIDS}}
    \caption{Warning message aimed to reduce anxiety}
    \label{fig:explainanxiety}
\end{figure*}

\end{document}